\documentclass[pra,twocolumn,amsmath,amssymb]{revtex4}

\usepackage{graphicx}
\usepackage{dcolumn}
\usepackage{float} 
\usepackage[charter]{mathdesign}
\usepackage{bm}

\usepackage[colorlinks = true,
            linkcolor = blue,
            urlcolor  = blue,
            citecolor = blue,
            anchorcolor = blue]{hyperref}

\newcommand{\etal}{\textit{et al.}} 
\newcommand\kv{\mathbf{k}}

\begin{document}

\title{Oxygen hole content, charge-transfer gap, covalency, and cuprate superconductivity} 

\author{N. Kowalski}
\affiliation{D\'epartement de physique and Institut quantique, Universit\'e de Sherbrooke, Sherbrooke, Qu\'ebec, Canada J1K 2R1}
\email{Nicolas.Kowalski@usherbrooke.ca}

\author{S. S. Dash}
\affiliation{D\'epartement de physique and Institut quantique, Universit\'e de Sherbrooke, Sherbrooke, Qu\'ebec, Canada J1K 2R1}
\email{Sidhartha.Shankar.Dash@USherbrooke.ca}

\author{D. S\'en\'echal}
\affiliation{D\'epartement de physique and Institut quantique, Universit\'e de Sherbrooke, Sherbrooke, Qu\'ebec, Canada J1K 2R1}

\author{A.-M. S. Tremblay}
\affiliation{D\'epartement de physique and Institut quantique, Universit\'e de Sherbrooke, Sherbrooke, Qu\'ebec, Canada J1K 2R1}
\date{\today}

\begin{abstract}
Experiments have shown that the families of cuprate superconductors that have the largest transition temperature at optimal doping also have the largest oxygen hole content at that doping. They have also shown that a large charge-transfer gap, a quantity accessible in the normal state, is detrimental to superconductivity. 
We solve the three-band Hubbard model with cellular dynamical mean-field theory and show that both of these observations follow from the model. Cuprates play a special role amongst doped charge-transfer insulators of transition metal oxides because copper has the largest covalent bonding with oxygen. 
\end{abstract}

\date{\today}


\maketitle

\section{Introduction}
Although several classes of high-temperature superconductors have been discovered, including pnictides~\cite{Kamihara_Watanabe_Hirano_Hosono_2008}, sulfur hydrides~\cite{Drozdov_Eremets_Troyan_Ksenofontov_Shylin_2015} and rare earth hydrides~\cite{Drozdov_Kong_Minkov_Besedin_2019,Kong_Minkov_Kuzovnikov_Besedin_Drozdov_2019}, cuprate high-temperature superconductors are still particularly interesting from a fundamental point of view because of the strong quantum effects expected from their doped charge-transfer insulator nature~\cite{Anderson:1987}, and single-band spin-one-half Fermi surface~\cite{Czyzyk_Sawatzky_1994,andersen_lda_1995}.

Amongst the most enduring mysteries of cuprate superconductivity is the experimental discovery, early on, that the hole content on oxygen plays a crucial role~\cite{tranquada_x-ray-absorption_1987,fujimori_spectroscopic_1987,andersen_lda_1995,Zheng_Kitaoka_Ishida_Asayama_1995}. Oxygen hole content ($2n_p$) is particularly relevant since NMR~\cite{rybicki_perspective_2016} suggests a correlation between $T_c$ and $2n_p$ on the CuO$_2$ planes: A higher oxygen hole content at the optimal doping of a given family of cuprates leads to a higher critical temperature. This is summarized in Fig.~2 of Ref.~\cite{rybicki_perspective_2016}. The charge transfer gap also seems to play a central role for the value of $T_c$, as suggested by scanning tunneling spectroscopy~\cite{ruan_relationship_2016} and by theory~\cite{weber_scaling_2012}. Many studies have shown that doped holes primarily occupy oxygen sites~\cite{tranquada_x-ray-absorption_1987,emery_mechanism_1988,chen_electronic_1991,gauquelin_atomic_2014}. It is this long unexplained role of oxygen hole content on the strength of superconductivity in cuprates that we address in this paper. 

The vast theoretical literature on the one-band Hubbard model
in the strong-correlation limit shows that many of the qualitative experimental features of cuprate superconductors~\cite{gull_numerical_2015,pavarini_emergent_2013} can be understood~\cite{Alloul_2014}, but obviously not the above experimental facts regarding oxygen hole content. Restricting ourselves to hole-doped cuprates and to a small sample of the literature on the one-band model, cluster extensions of dynamical mean-field theory~\cite{KotliarRMP:2006,Maier:2005,LTP:2006} lead to a phase diagram with antiferromagnetism and $d$-wave superconductivity~\cite{Maier:2000a,lichtenstein2000antiferromagnetism, maier_d:2005,
kancharla2008anomalous,
haule2007strongly,gull_superconductivity_2013,fratino2016organizing,Foley_Verret_Tremblay_Senechal_2019}, a Knight shift pseudogap temperature exhibiting the correct temperature and doping dependence~\cite{Senechal:2004,Ferrero:2009,GullFerrero:2010,Imada:2011,Sordi_Semon_Haule_Tremblay_2012,fratino_pseudogap_2016,Wu_Ferrero_2018,Scheurer_Chatterjee_Wu_Ferrero_Georges_Sachdev_2018,Maier_Scalapino_2019,LeBlanc_Li_Chen_Levy_Antipov_Millis_Gull_2019,Reymbaut:2019}, and even phase fluctuations in the underdoped regime~\cite{Maier_Scalapino_2019,Simard_Hebert_Foley_Senechal_Tremblay_2019} as suggested by phenomenology~\cite{emery_importance_1995-1} and by the Uemura relation~\cite{uemura_universal_1989}. However, variational calculations~\cite{corboz_competing_2014} and various Monte Carlo approaches~\cite{Aimi:2007,Huang_Mendl_Liu_Johnston_Jiang_Moritz_Devereaux_2017} suggest that $d$-wave superconductivity in the one-band Hubbard model may not be the ground state, at least in certain parameter ranges~\cite{fradkin_colloquium_2015,Qin_Chung_Shi_Vitali_Hubig_Schollwock_White_Zhang_2020}. 

It is thus important to investigate more realistic models, such as the three-band Emery-VSA model that accounts for copper-oxygen hybridization of the single band that crosses the Fermi surface~\cite{emery_theory_1987,Varma_Schmitt_Rink_Abrahams_1987}. A variety of theoretical methods~\cite{de2009correlation,wang_role_2011,go_spatial_2013,weber_scaling_2012,fratino_pseudogap_2016,Teranishi_Nishiguchi_Kusakabe_2020,Zegrodnik_Biborski_Fidrysiak_Spalek_2020,cui_ground-state_2020,Mai_Balduzzi_Johnston_Maier_2021} revealed many similarities with the one-band Hubbard model, but also differences related to the role of oxygen~\cite{peets:2009,Weber:2014,Ebrahimnejad_Sawatzky_Berciu_2016}. 

Investigating the role of oxygen and of the charge-transfer gap
on the relative value of the transition temperature $T_c$ for various cuprates is a key scientific goal of the quantum materials roadmap~\cite{Giustino_Lee_Trier_Bibes_Winter_Valenti_2021} ~\footnote{A few phenomenological correlations have been found to control the optimal $T_c$ for a given family of cuprates. ``Homes law''~\cite{Homes_Dordevic_2004} and ``Basov's law''~\cite{
    dordevic2013organic} focus on the behavior of the normal state conductivity.}. We do find and explain the above correlations found in NMR and in scanning tunnelling spectroscopy, highlight the importance of the difference between electron affinity of oxygen and ionization energy of copper~\cite{Varma_Schmitt_Rink_Abrahams_1987,Varma_LesHouches}. 

We do not address questions related to intra unit-cell order~\cite{Mukhopadhyay_Sharma_Kim_Edkins_Hamidian_Eisaki_Uchida_Kim_Lawler_Mackenzie_2019,Varma_review_2019}.

\section{Model and Method}
In second-quantized notation, the three-band Emery-VSA Hubbard model~\cite{emery_theory_1987,Varma_Schmitt_Rink_Abrahams_1987,andersen_lda_1995} on the square lattice is 
\begin{equation}
H = \sum_{\kv \sigma} \Psi^{\dagger}_{\kv \sigma}\mathbf{h}_0(\kv )\Psi_{\kv \sigma} + U\sum_i n^d_{i\uparrow} n^d_{i \downarrow}
\end{equation}
where the multiplet $\Psi^{\dagger}_{\kv \sigma}=$($d^{\dagger}_{\kv ,\sigma}$, $p^{x\dagger}_{\kv ,\sigma}$, $p^{y\dagger}_{\kv ,\sigma}$) contains the creation operators for electrons on the copper $d_{x^2-y^2}$ and the oxygen $p_x$ and $p_y$ orbitals ($\kv$ is the wave-vector and $\sigma$ the spin projection) and $n^d_{i\sigma} = d^{\dagger}_{i\sigma} d_{i\sigma}$. Taking the distance between unit cells to be unity, the non-interacting Hamiltonian $ \mathbf{h}_0(\kv )$ is given in Eq.~\eqref{eq:h0}.
\begin{figure*}[bt!]
\begin{equation}\label{eq:h0}
\mathbf{h}_0(\kv ) = 
\begin{pmatrix}
\epsilon_d & t_{pd}(1 - e^{-ik_x}) & t_{pd}(1 - e^{-ik_y})\\
t_{pd}(1 - e^{ik_x}) & \epsilon_p + 2t_{pp}'\cos k_x & t_{pp}(1 - e^{ik_x})(1 - e^{-ik_y}) \\
t_{pd}(1 - e^{ik_y}) & t_{pp}(1 - e^{-ik_x})(1 - e^{ik_y}) & \epsilon_p + 2t_{pp}'\cos k_y
\end{pmatrix}
\end{equation}
\end{figure*}

\begin{figure*}
\includegraphics[width=1\textwidth]{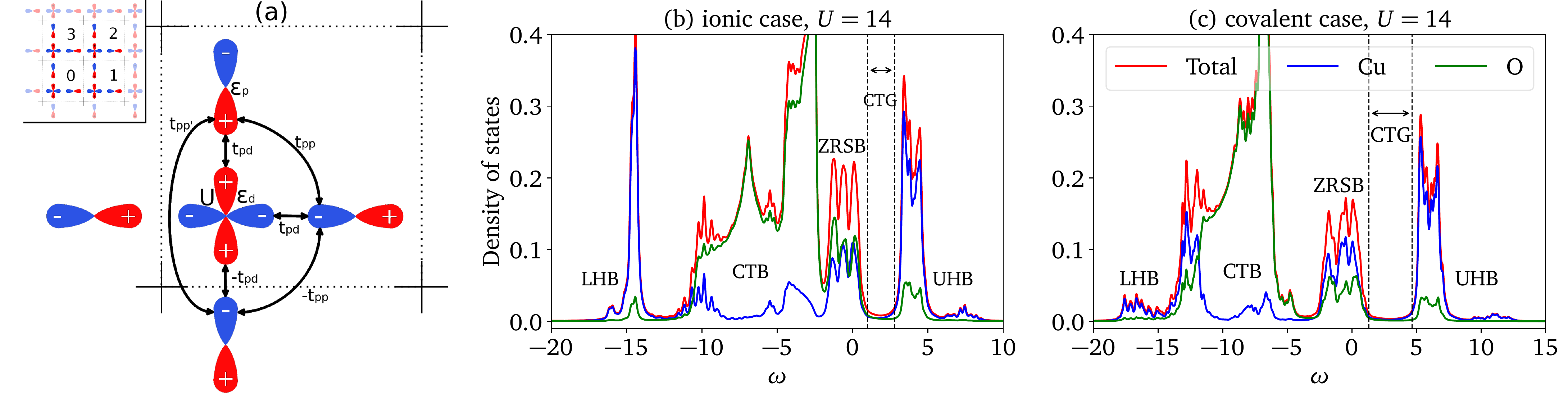}
\caption{\label{fig:1} 
a) Schematic view of the three-band Hubbard model. The $d$-shaped orbitals sit on copper atoms whereas the two types of $p$-shaped orbitals ($p_x$ and $p_y$) sit on the surrounding oxygen atoms. Inset: the $2\times2$ cluster used for the CDMFT calculation. b) Total and partial densities of states of the interacting model for parameter set \eqref{eq:fratino_params} (ionic case) for 12\% hole doping at $T=0$. c) The same, for parameter set \eqref{eq:Bi2212_params} (covalent case) for 13\% hole doping at $T=0$. LHB stands for lower Hubbard band, UHB for upper Hubbard band, CTG for charge-transfer gap, CTB for charge-transfer band and ZRSB for Zhang-Rice singlet band. Note that the isolated peak in the lower Hubbard band appears clearly only for the ionic case Eq.~\eqref{eq:fratino_params}.
}
\end{figure*}

The on-site energies on the Cu and O orbitals are noted $\epsilon_d$ and $\epsilon_p$, respectively. 
We chose not to include the $-2t_{pp}$ contribution to the on-site oxygen energy. 
We can set $\epsilon_d = 0$ without loss of generality. 
The first-neighbor Cu-O hopping is $t_{pd}$, the first neighbor (diagonal) O-O hopping is $t_{pp}$ and O-O hopping through 
the Cu site is $t_{pp}'$. 
Finally we only consider interactions on the Copper sites ($U$). 
This is justified by DFT+U calculations showing the on-site oxygen and inter-site interactions to be much smaller than the on-site Copper interaction~\cite{mcmahan_calculated_1988,hybertsen_calculation_1989}. 
The Hamiltonian and the unit cell is visualized on Fig.~\ref{fig:1}(a). 

We use cellular dynamical mean field theory (CDMFT)~\cite{KotliarRMP:2006} on a cluster of four unit cells (inset of Fig.~\ref{fig:1}(a)) in order to capture the local fluctuations, crucial for $d$-wave superconductivity. The environment is a bath of non-interacting electrons that hybridizes self-consistently with the cluster. 
Since the eight oxygen atoms on the cluster are uncorrelated, they may be integrated out and incorporated as a constant hybridization function in the lattice Green function that is needed for the CDFMT self-consistency relation. We can then concentrate on a four-site copper cluster that is solved with exact diagonalization with eight bath sites at $T=0$~\cite{dash_pseudogap_2019}. Because there is no direct hopping between copper sites, at finite $T$ the cluster in an infinite bath problem can be solved using the continuous-time quantum Monte Carlo algorithm based on the segment algorithm~\cite{WernerMillis:2006} as in Refs. ~\cite{fratino_pseudogap_2016,gull_continuous-time_2011}.

The three-band Hubbard model describes a charge-transfer insulator at large values of $U$ and at a filling of five electrons per unit cell (the undoped state). 
Such an insulating state is realized as $U$ increases and splits the Cu band into lower and upper Hubbard bands, such that the upper band is pushed beyond the oxygen-dominant band (around $\epsilon_p$), leading to an insulating gap between the two. 
In the strong-coupling limit, excitations from the undoped state requires a transfer of charge from the O orbital to the Cu orbital; hence the insulating gap is referred to as a charge-transfer gap (CTG) and the insulating state as a charge-transfer insulator. 
The central band is further called the charge-transfer band (CTB). 
This is different from a Mott insulator (e.g. in the one-band Hubbard model) where the insulating (or Mott) gap appears between the lower and upper Hubbard bands.
On doping the charge-transfer insulator, the holes primarily go into the oxygen orbitals and another band appears at the Fermi level referred to as the Zhang-Rice singlet band (ZRSB)~\cite{unger1993spectral,chen2013doping,Mai_Balduzzi_Johnston_Maier_2021}. 
Zhang-Rice singlets are characterized by singlet-states formed between the Cu orbital and the adjacent O orbitals~\cite{zhang1988effective,brookes2015stability}. 

Fig.~\ref{fig:1}(b),(c) shows the density of states for the following two sets of parameters (in units of $t_{pp}\sim 0.65 {\rm eV}$), using as a zero of energy the copper site energy $\epsilon_d=0$:
\begin{align}
 \label{eq:fratino_params}\epsilon_p &= 7 &t_{pd} &= 1.5 &t_{pp}' &= 1 \\
 \label{eq:Bi2212_params}\epsilon_p &= 2.3 &t_{pd} &= 2.1 &t_{pp}' &= 0.2.
\end{align}
In these units, $\epsilon_p$ measures the difference between the oxygen and copper site energies. Since $\epsilon_p$ is large for the case in Eq.~\eqref{eq:fratino_params}, for conveninece we refer to it as the ionic case and since $\epsilon_p$ is small for the case in Eq.~\eqref{eq:Bi2212_params}, we refer to it as the covalent case. 
In Eq.~\eqref{eq:Bi2212_params} the value of $\epsilon_p$ has been obtained following the estimates for Bi2212 in Weber~\etal~\cite{weber_scaling_2012}, subtracting the double counting contribution. The parameters in Eq.~\eqref{eq:fratino_params} are obtained from Ref.~\cite{fratino_pseudogap_2016}. 
As seen from the density of states (Fig.~\ref{fig:1}(b), (c)), the main difference between the two parameter sets is the absence of a distinct lower Hubbard band (LHB) for the covalent case Eq.~\eqref{eq:Bi2212_params} compared to the ionic case Eq.~\eqref{eq:fratino_params} . This is due to a much smaller value of $\epsilon_p$ in the former, leading to better co-valency and to a mixing of the lower Hubbard band and the 
charge-transfer band. This is a more realistic description of cuprates, whereas the ionic case Eq.~\eqref{eq:fratino_params} describes a scenario with well-separated lower Hubbard band and charge-transfer band. 
Also, the covalent case Eq.~\eqref{eq:Bi2212_params} has a well formed Zhang-Rice singlet band compared to the ionic case Eq.~\eqref{eq:fratino_params}. 
Finite-temperature calculations were not possible for the parameters of the covalent case because of the sign problem.

\section{Results}

We define the order parameter $\psi =\langle \hat\Delta\rangle/N$, where $N$ is the total number of unit cells in the lattice and $\hat\Delta$ is the pairing operator:
\begin{equation}
\hat\Delta = \sum_{\langle ij\rangle_x}\left(d_{i,\uparrow}d_{j,\downarrow}-d_{i,\downarrow}d_{j,\uparrow}\right)\\
-\sum_{\langle ij\rangle_y}\left(d_{i,\uparrow}d_{j,\downarrow}-d_{i,\downarrow}d_{j,\uparrow}\right) + \mathrm{H.c.}
\end{equation}
where $\langle ij\rangle_x$ and $\langle ij\rangle_y$ indicate nearest-neighbor copper sites in the $\hat{x}$ and $\hat{y}$ directions respectively.

\begin{figure*}
\includegraphics[width=1\textwidth]{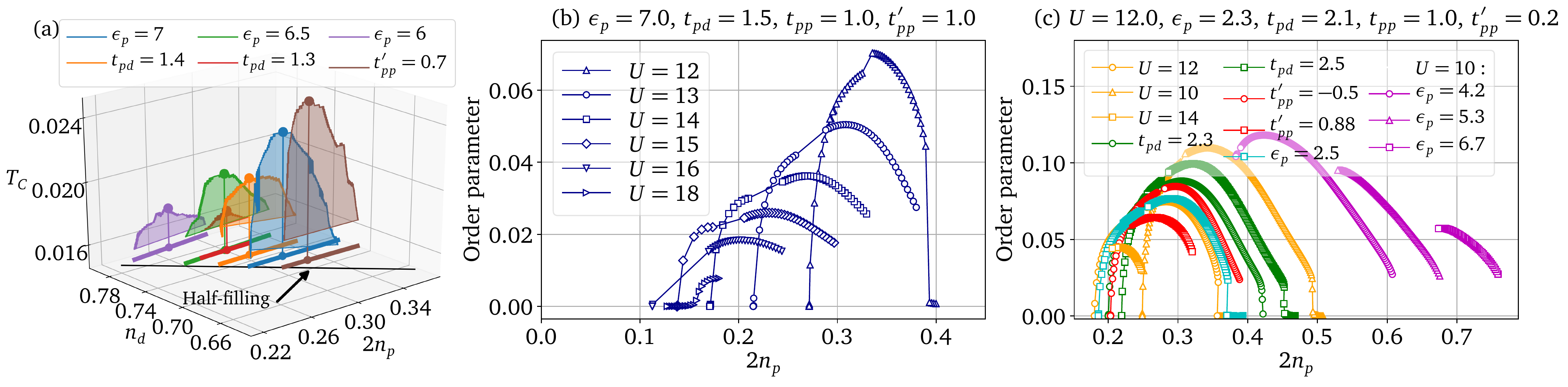}
\caption{\label{fig:superconducting-domes} 
Superconducting domes as a function of atomic hole contents. a) critical temperature as a function of oxygen $2n_p$ and copper $n_d$ hole contents for various parameter sets obtained by varying one parameter at a time, starting from the ionic case Eq.~\eqref{eq:fratino_params} at $U=12$. The solid black line $n_d+2n_p=1$ corresponds to the parent compound. b), c) Superconducting order parameter at $T=0$ as a function of oxygen hole content ($2n_p$), for various parameter sets that deviate slightly from, respectively, the ionic \eqref{eq:fratino_params} and the covalent~\eqref{eq:Bi2212_params} cases with parameters indicated in the legends. %
}
\end{figure*}

As we aim to relate a change in oxygen occupation to a change in optimal $T_c$, we need to obtain multiple superconducting domes, corresponding to different parameter sets, and compare them. This is shown on Fig.~\ref{fig:superconducting-domes}(a) as a function of oxygen and copper hole contents for parameters in the vicinity of the ionic case. The presentation follows that of the experimental paper of Ref.~\cite{rybicki_perspective_2016}. 
There is a correlation between the optimal $T_c$ (vertical lines) and $2n_p$.
Increasing $2n_p$ leads in general to a higher $T_c$. 
However this correlation is not absolute: for instance, the green and red domes at around $n_d\approx0.76$ overlap without having the same maximum $T_c$.

We also show the order parameter at zero temperature as a function of $2n_p$ in Fig.~\ref{fig:superconducting-domes}(b,c) for variations of the parameters in the vicinity, respectively, of the ionic case Eq.~\eqref{eq:fratino_params} and of the covalent case Eq.~\eqref{eq:Bi2212_params}. For the covalent case, we studied a range of parameters that corresponds to those seen in cuprates. 
The maximum superconducting order parameter decreases with $U$, as in the one-band case~\cite{ofer2006magnetic,fratino2016organizing,kancharla2008anomalous,haule2007strongly}. 
We also see a positive correlation in the height of the superconducting domes (the maximum order parameter) with optimal $2n_p$, except beyond a certain $2n_p$ where the maximum order parameter decreases. As we now discuss, this corresponds to the closing of the charge transfer gap, a case not encountered in the cuprates. 

\begin{figure*}
\includegraphics[width=1\textwidth]{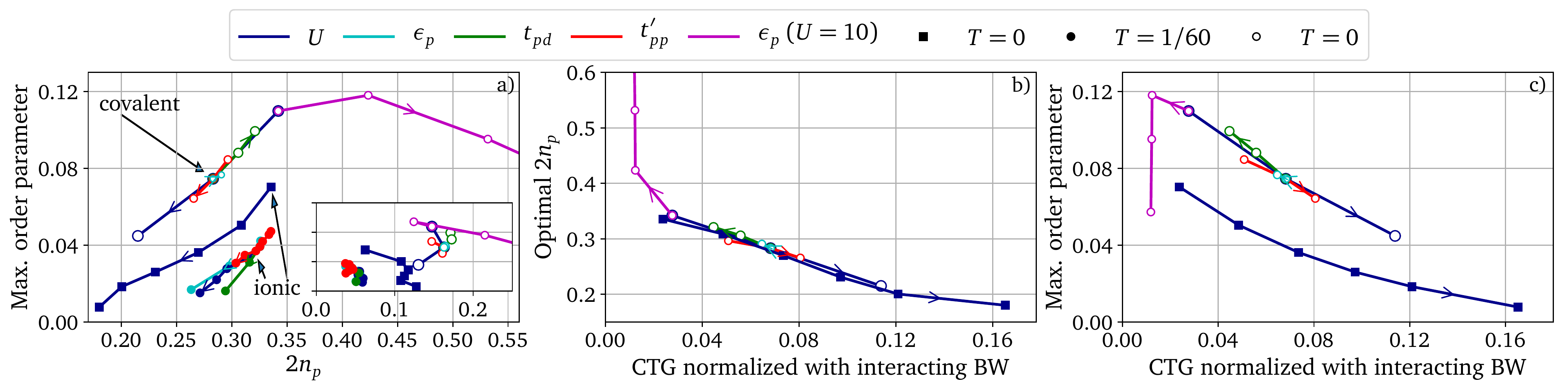}
\caption{ 
a) Maximum order parameter for each of the superconducting domes in Fig.~\ref{fig:superconducting-domes} as a function of the corresponding oxygen hole content ($2n_p$). The open symbols correspond to variations from the covalent case Eq.~(\ref{eq:Bi2212_params}) and the closed symbols correspond to variations from the ionic case Eq.~(\ref{eq:fratino_params}). The four curves with smallest order parameters and filled circles were obtained at $\beta=60$. 
We indicate the points corresponding to the reference ionic and covalent parameters with dark arrows.
Inset: maximum order parameter as a function of total doping. 
b) Oxygen hole content ($2n_p$) versus normalized CTG at optimal doping.
We have normalized the CTG with the total bandwidth to compare across different parameter sets.
c) Maximum order parameter as a function of the normalized CTG at optimal doping.  
For each color, only one parameter is changed. The arrows on the colored segments of the plots point towards an increase of the respective parameters indicated in the legend.
}\label{fig:correlations}
\end{figure*}

In order to see this trend more clearly and to compare all the data presented thus far, we plot in Fig.~\ref{fig:correlations}(a) the maximum order parameter that can be obtained at the top of the dome for a given set of microscopic parameters, and we group the results by color according to the model parameter that is varied. The black arrows mark the reference parameters for the ionic Eq.~\eqref{eq:fratino_params} and covalent Eq.~\eqref{eq:Bi2212_params} cases respectively, for $U=12$.

The dark blue curve (closed squares) and the curves below it are for parameters in the vicinity of the ionic case and the other curves above are for parameters in the vicinity of the covalent case. The four lowest curves were obtained at finite temperature and are therefore lower than the one obtained at $T=0$.
If we set aside the purple curve that corresponds to the closing of the charge-transfer gap, the correlation between the maximum order parameter and $2n_p$ still applies locally across all parameter sets. And the slope of all curves are similar. It is quite striking that the covalent case, corresponding more closely to parameters actually encountered in cuprates, has maximal values of the order parameter at given oxygen hole content that are larger than for the ionic case. 

The inset of Fig.~\ref{fig:correlations}(a) shows a very important result. There, we plot the same achievable maximum order parameter this time as a function of {\it total} doping, or hole content, as is usually done in plots of the phase diagram. 
We clearly notice the absence of correlation between the maximum order parameter and {\it total} doping, in stark contrast with what we see in the plot of the main panel of Fig.~\ref{fig:correlations}(a) as a function of oxygen hole content.

Finally, following experimental~\cite{ruan_relationship_2016} and theoretical~\cite{weber_scaling_2012} work linking the CTG to the critical temperature, we inspect the relation of the CTG, accurately accessible at $T=0$, to both oxygen occupation and order parameter. 
Fig.~\ref{fig:correlations}(b) shows the oxygen occupation as a function of the CTG, normalized to the bandwidth; the two quantities are clearly correlated.
We can explain this trend very easily by turning to the density of states. 
Indeed, when increasing the CTG, we reduce the overlap of the oxygen spectrum with the upper Hubbard band
This reduces the oxygen weight in the upper Hubbard band leading to a smaller $2n_p$.

Furthermore, we note from Fig.~\ref{fig:correlations}(c) that the order parameter decreases monotonously as the CTG increases, 
except for very low values of the CTG, as seen also in Ref.~\cite{Mai_Balduzzi_Johnston_Maier_bis_2021}; this is consistent with the experimental observation~\cite{ruan_relationship_2016} that the maximum $T_c$ decreases as the CTG increases. 
Once the CTG has opened, increasing the CTG reduces both the order parameter and the oxygen hole content ($2n_p$), hence making them behave monotonously. 

Fig.~\ref{fig:correlations}(c) clearly shows that the order parameter is larger for the model parameters near the covalent case 
Eq.~\eqref{eq:Bi2212_params} compared to the ionic case Eq.~\eqref{eq:fratino_params}, demonstrating that a lower value of 
$\epsilon_p - \epsilon_d$ favors superconductivity, as also seen in Ref.~\cite{weber_scaling_2012} and suggested early on in 
Ref.~\cite{Varma_Schmitt_Rink_Abrahams_1987,Varma_LesHouches}. 

\section{Discussion}\label{Sec:Discussion}

The range of parameters that we have explored for the covalent case, while relatively small, is sufficient to cover 
the range of parameters that corresponds to existing families of cuprate high-temperature 
superconductors~\cite{weber_scaling_2012}. The ionic case corresponds to a large change in model 
parameters and it is not a realistic model of cuprate superconductivity. We also studied the unphysical 
limit where the CTG disappears to establish that it is crucial for superconductivity. 

Different parameter sets correspond to different compounds, or the same compound in different physical situations. 
For example, $\epsilon_p$ ($\epsilon_d=0$) is strongly influenced by the presence and location 
of the apical oxygen~\cite{Weber_Haule_Kotliar_2010}. Also, applying a positive pressure on 
compounds clearly increases the hopping parameters. 

The percentage change in each parameter that is necessary for a 1\% relative change in the optimal 
superconducting order parameter is shown in Table~\ref{ParameterChange} for the two cases studied here. 
The reference values are taken to be the calculations done with Eq.~\eqref{eq:fratino_params} (ionic case) and 
Eq.~\eqref{eq:Bi2212_params} (covalent case) at $U=12$. Clearly, the percentage change of 
parameters that is necessary to increase the optimal superconducting order parameter is larger 
for the covalent case than for the ionic case. Since the reference values of the optimal order parameter of both cases are close, this suggests that the parameters for the covalent 
case are closer to the maximum achievable value of the order parameter in the parameter space of the three-band 
model. 
Interaction on copper, $U$, has the largest impact on the superconducting order parameter
while oxygen-copper hopping $t_{pd}$ is the second most important parameter. 

\begin{table} 
\caption{\label{ParameterChange} Percentage change of each parameter that is necessary to 
cause a 1\% relative increase in the optimal superconducting order parameter (SC) and in the optimal CTG. 
Reference model parameters are Eq.~\eqref{eq:fratino_params} for the ionic case and 
Eq.~\eqref{eq:Bi2212_params} for the covalent case
with $U=12$ for both cases.} 

\def\arraystretch{1.5}
\medskip
\centering
\begin{tabular}{lD..2D..2D..2} 
\hline\hline
Model: & \multicolumn{2}{c}{covalent} & \multicolumn{1}{c}{ionic} \\
& \multicolumn{1}{c}{SC} & \multicolumn{1}{c}{CTG} & \multicolumn{1}{c}{SC} \\
\hline
$\Delta U/U$ (\%) & -0.38 & 0.24 & -0.14 \\
$\Delta t_{pd}/t_{pd}$ (\%) & 0.57 & -0.68 & 0.23 \\
$\Delta \epsilon_p/\epsilon_p$ (\%) & 3.21 & -1.65 & 0.25 \\
$\Delta t_{pp}/t_{pp}$ (\%) & 1.66 & - 0.48 & -2.5 \\
$\Delta t'_{pp}/t'_{pp}$ (\%) & -25.38 & 15.16 & -1.4 \\
\hline
\end{tabular}

\end{table}

Given the strong correlation that we found between oxygen hole content at optimal doping and charge-transfer gap (CTG), either quantity can be optimized. Recall however that oxygen hole content increases the maximum order parameter while the CTG decreases it. This is illustrated in Table~\ref{ParameterChange} for the covalent case. 

As we saw, covalency, that is mostly controlled by the difference between the ionization energy of the transition metal 
and the oxygen affinity~\cite{Varma_Schmitt_Rink_Abrahams_1987,Varma_LesHouches}, is crucial for 
high-temperature superconductivity. Among $3d$ transition metals, copper forms the most covalent 
bonds with oxygen (also favored due to the two-dimensional structure), hence other transition metal oxides are 
less likely to be high-temperature superconductors. Nickelates~\cite{Li_Lee_Wang_Osada_Crossley_Lee_Cui_Hikita_Hwang_2019}, 
for example, are more ionic and have a lower $T_c$. This suggests that combining other transition metal 
oxides with other chalcogens or pnictides could be a better way to look for compounds with stronger superconductivity. 

\section{Conclusion}
First, we have shown that the experimental correlation between 
oxygen hole content ($2n_p$) at optimal doping and 
optimal $T_c$ is satisfied in both the ionic and covalent limits of the three-band model, 
thus resolving a long-standing theoretical issue. In each case (covalent or ionic), changes 
in different model parameters that lead to the same $2n_p$ lead, with few exceptions, 
to the same superconducting order parameter. By contrast, total doping can lead to different maximum order parameters, depending on model parameters.

Second, we have also shown that $2n_p$ and the size 
of the charge-transfer gap (CTG) are almost perfectly correlated. So oxygen hole content, or equivalently the
CTG, are the important parameters that control superconductivity. Since the CTG should be nearly 
independent of temperature starting from room temperature to around 100~K, measuring the charge transfer gap in the 
normal state should give a strong indication of whether or not a specific compound can be a 
high-temperature superconductor. A small CTG is needed for superconductivity to appear but a large 
CTG is detrimental. 

Finally, comparing the results for the ionic and covalent cases, we also conclude that 
cuprates play a special role amongst doped charge-transfer insulators of transition metal oxides 
because copper has the largest covalent bonding with oxygen. 

\begin{acknowledgments}
We thank H. Alloul, J. Haase, G. Kotliar, Peizhi Mai, G. Sordi, T. Maier, C.M. Varma and P. Werner for useful discussions. This work has been supported by the Natural Sciences and Engineering Research Council of Canada (NSERC) under grants RGPIN-2019-05312 and RGPIN-2020-05060, and by the Canada First Research Excellence Fund. Simulations were performed on computers provided by the Canadian Foundation for Innovation, the Minist\`ere de l'\'Education des Loisirs et du Sport (Qu\'ebec), Calcul Qu\'ebec, and Compute Canada.
\end{acknowledgments}


\end{document}